\documentstyle[12pt]{article}
\begin {document}
\begin{flushright}
MRC-PH.TH-13/96   \\

gr-qc/9611015

\end{flushright}
\vspace{1.0truein}

\centerline{\bf{ An Extension of 
Multiple Cosmic String Solution: A Proposal}}
\vspace{1.0truein}
\centerline{M.Horta\c csu $^{\dagger *}$ and N. \" Ozdemir $ {\dagger} $} 
\vspace{.8truein}
\centerline{ ${\dagger}$ Physics Department, Faculty of Sciences and Letters}
\vspace{.3truein}
\centerline { ITU 80626, Maslak, Istanbul, Turkey}
\vspace{.6truein}
\centerline {${*}$ Physics Department, 
T\" UBITAK Research Institute for Basic Sciences, Turkey}
\vspace{2.5truein}
\noindent 
{\raggedright{\bf{Abstract}}
We extend the work done for cosmic strings and show that for a more
general class of locally flat metrics 
the one loop calculation do not introduce
 any new divergences to the VEV of the energy of a scalar particle.  We
 explicitly perform the calculation for the configuration 
 where we have one cosmic
 string in the presence of a dipole made out of cosmic strings.}\\
 \vspace{.5truein}
 PACS number is 98.80 Cq

 \vfill\eject
 \newpage
 \baselineskip=18pt

 \noindent
Extensive work has been done on cosmic strings $^{/1}$ and other
topological defects.  A remarkable property of the field theoretical
 calculations using cosmic strings as background is the existence of a finite
 contribution to the energy of a scalar particle due to the presence 
 of the cosmic string $^{/2} $.  In a beautiful paper  $^{/3}$ A.N. Aliev
 showed that there is an intricate cancellation 
 mechanism in the cosmic string
 background which makes the perturbative expression finite.  The divergence
  due to the one loop integration is cancelled by the zero coming from the
  anti Fourier transforming the resulting expression.

In this note we want to point out to a simple property of certain metrics,
made out of two " holomorphic " 
structures, where  similar cancellation occurs.
In this sense we extend the 
Aliev result to more general , still locally flat,
metrics.

 Take the special form of the Dirac operator
 $$ D'= e^{-g_1(x_0-\gamma_0 \gamma_3 x_3)} 
 (\gamma_0 \partial ^0 + \gamma_3 \partial ^3)
 +e^{-g_2 (x_1-\gamma_1 \gamma_2 x_2)} 
 (\gamma_1 \partial ^1 +\partial_2 \partial ^2). \eqno {1} $$
 Here $\gamma^{\mu} $ are Dirac gamma matrices 
 obeying the anticommutation relations
 $[\gamma^{\mu},\gamma^{\nu} ]_{+} = 2g^{\mu \nu} $, and $g_1,g_2$ are
 arbitrary smooth functions which are defined by their power expansions.
 Define the d'Alembertian operator as
 $$ \Box =  {Tr D'D'\over {4}}. \eqno {2} $$
 One sees that the d'Alembertian operator reads
 $$ \Box = e^{-f_1(x_0,x_3)}\left({{\partial^{2}} \over {\partial x_0^2}}
 -{{\partial^{2}} \over {\partial x_3^2}}\right)-
 e^{-f_2(x_1,x_2)} \left({{\partial^2} \over {\partial x_1^2}}
 +{{\partial^2} \over {\partial x_2^2}} \right) ,\eqno {3} $$
 where
 $$ f_1(x_0,x_1)= {Tr \over {4}} \left(g_1(x_0+\gamma_0 \gamma_3 x_3) +
 g_1(x_0-\gamma_0 \gamma_3 x_3)\right) \eqno {4a} $$
 $$ f_2(x_1,x_2)= {Tr \over {4} } \left(g_2(x_1+\gamma_1 \gamma_2 x_2)+
 g_2(x_1- \gamma_1 \gamma_2 x_2) \right) .\eqno {4b} $$
 
One can derive the metric that will give this d'Alembertian operator,
using the relation  $ \Box =\frac{1}{ \sqrt{-g}}\partial_\mu
g^{\mu \nu} \sqrt{-g} \partial_\nu $  .
This d'Alembertian can be derived from the metric
$$ds^2= e^{-f_1(x_0,x_3)} (dx_0^2 -dx_3^2) -e^{-f_2(x_1,x_2)} (dx_1^2+dx_2^2) \eqno {5} $$
where $f_1,f_2$ are defined as in eq.(4).

 Note that the metrics given in equation (5) are all locally flat, vacuum
 solutions, allowing possible Dirac delta function type singularities.

 From this point on we will restrict our 
 discussion to the metric where $f_1=1$
 and $f_2$ an arbitrary smooth function.  The more general case given above 
 is twice this structure.  For this case the Dirac operator in the
 general case can be derived using standard procedure.
 $$ D= \gamma_0 \partial^0 +\gamma_3 \partial ^3+ e^{{{f(x_1,x_2)} \over {2}}}
 [\gamma _1 \partial ^1+ \gamma_2 \partial ^2 + {{\gamma_1} \over {4}} 
 {{\partial_1 f } }
 +{{\gamma_2} \over {4}} {{\partial_2 f} } ] .\eqno {6} $$
 For this general case, the square of the Dirac operator does not give us
 the d'Alembertian.  There are extra terms proportional to derivatives
 of $f_2$.  We state this fact by writing
 $$ \Box \not= {Tr\over{4}} DD. \eqno {7} $$
 If we choose $f$so that it can be written as 
 $f=g_2(x_1-\gamma_1\gamma_2 x_2)$ , we get 
 $${{\gamma_1} \over {4}} {{\partial g_2} \over {\partial x_1}}
 +{{\gamma_2} \over {4}} {{\partial g_2} 
 \over {\partial x_2}}\equiv 0 .  \eqno {8} $$
 Then $D'=D$ and eq. (5) is satisfied with $D$ as well for this special
 case.   To derive this result
 we used the series expansion for $g_1$ and the relation
 $\gamma_1 \gamma_2 \gamma_1 \gamma_2 =-1$.

 With this choice for $f_1$ and $f_2$ the Dirac operator 
 is the true square root 
 of the d'Alembertian, similar to its behaviour 
 in the free case.  Our examples 
 are extensions of the free case, since our metrics are all locally 
 flat.  One should recall that solutions with only topological defects also
 have this property. This gives us the possibility of finding 
 interesting solutions which generalize
 the cosmic string solutions.
 Below, using semi-classical methods, we will study two cases where   
 the scalar 
 field is coupled to the metric 
  .

   If we do not want to extend  our calculations to spinors, we can replace
   $\gamma_1 \gamma_2$ by $i=\sqrt{-1}$ and $\gamma_0 \gamma_3$ by unity,
   where both are multiplied by the unit matrix.  
   We will use the latter expressions
   here, since we will be confined only to calculations 
   with the scalar field
   in this paper.

   A.N.Aliev $^{/3  }$  has shown that at first 
   order perturbation theory, the cosmic
   string gives a  finite contribution to the vacuum energy.  This
   result is in accordance with the expectations, since it is well known
   that even the exact result does not introduce additional infinities.
   Our result is the extension of the Aliev result. The multiple cosmic
   string solution is a special case of our case since we can write
   $$  \sum_{i=1} ^n \log (x_{1i} ^{2} + x_{2i} ^{2} 
   ) = {Tr \over {4}} \left(
   \sum_{i=1} ^{n} (\log ( x_{i1} +\gamma^1 \gamma^2 x_{i2} )+
   \log (x_{i1} - \gamma^1 \gamma^2 x_{i2} )) \right), \eqno{9} $$
   where the LHS of the equation corresponds to the multiple-cosmic string
   solution $^{/4  }$.  

   We can show that the similar finite result in first order perturbation 
   theory  can be obtained if $f_1, f_2 $ are taken in the form given in
   eq. (2).  
   The metric is locally flat , then, since we can define new coordinates, 
   $$ d \tau = e^{g_1(x_0+ x_3)} {{(dx_0+ dx_3)} 
   \over {\sqrt{2}}} \eqno  {10} $$
         $$ d \overline \tau = e^{g_1(x_0- x_3)} {{(dx_0-dx_3)} 
         \over {\sqrt{2}}} \eqno {11} $$
           $$ d \zeta= e^{g_2(x_1+i x_2)} {{ (dx_1+i dx_2)} 
           \over {\sqrt{2}}} \eqno  {12} $$
           $$ d \overline \zeta = e^{g_2(x_1-i x_2)} {{ (dx_1-i dx_2)} 
           \over {\sqrt{2}}} .\eqno {13} $$
   Then 
   $$ds^2=d\tau d\overline \tau- d\zeta d\overline \zeta ,\eqno {14} $$
   a metric which is locally flat.  Only zeros and singularities of 
   $g_1$ and $g_2$
   can introduce curvature at certain points, giving rise to 
   Dirac delta function
   type singularities.  For particular choices of $g_1$ and $g_2$, we see
   that to perform the transformation from $\zeta$ 
   and $\tau$ back to $x_1+ix_0$ and
   $x_0+x_3$     ,  we have to introduce cuts to the $\tau$ 
   and $\zeta$ planes.

   We will use this formalism to calculate the vacuum fluctuations 
   for one special forms
   , which can be interpreted as cosmic strings with positive 
   and negative masses.

\bigskip

As an example we will study the case where
$$g_2 = -\beta \log {(1+\alpha (z-{1\over{z}}))} \eqno{15} $$
When we study only the scalar particle case, we can take $z=x_1+ix_2$.  If
we want to extend our problem also to the spinor case, we replace $z$ by
$\xi=x_1+\gamma_1 \gamma_2 x_2$.
    and use the $Trace$ operation at appropriate points.
\noindent
We write
$$ \Box = \partial_0^2 -\partial_3^2-4 e^{\beta \log
(1+\alpha(z-{1\over {z}}))
+
\beta  \log (1+\alpha(\overline z -{1\over {\overline z}}))} 
{{\partial} \over {\partial \overline z}} {{\partial } \over 
{ \partial  z}} \eqno {16} $$
In first order perturbation theory we first expand the exponential 
and then the logarithm.
We end up with
$$ \Box_1 = \partial_0^2-\partial_1^2 -\partial_2^2 - 
\partial_3^2 -\beta \alpha
(2x_1-{{2x_1} \over {x_1^2+x_2^2}})(\partial_1^2 +\partial_2^2). 
\eqno {17} $$
The first order Greens Function reads
$$G_F^1(x-y) = \int dw G_0(x-w) V(w) G_0(w-y) \eqno {18} $$
where
$$ V=V_1(x_1,x_2)(\partial_1^2+\partial_2^2) \eqno {19} $$ and  
$$ V_1(x_1,x_2)=-\beta \alpha (2x_1 -{2x_1 \over {x_1^2+x_2^2}})  
. \eqno {20} $$
If we go to momentum space we have to calculate
$$ \int dq \int dp {{p_1^2+p_2^2} \over {p^2(p-q)^2}} e^{iqx} V(q) $$
to obtain the $G_F^1$ at the coincidence limit .To get $<T^{00}>$, the VEV
of the energy, we have to differentiate $G_F^1$.  This operation can be
shown to result in the integral
$$ \int dq \int dp {{(p_1^2+p_2^2)^2}\over {p^2(p-q)^2}} e^{iqx} V(q) $$
where
$$V(q)=\delta(q_0) \delta (q_3) \int dx_1 dx_2 V_1(x_1,x_2) 
e^{i(q_1 x_1+q_2 x_2)} $$
$V_1(x_1,x_2)$  was defined in eq. (20).

As Aliev has shown $^{/3}$ this calculation boils down to 
multiplying $V_1(q)$ by
$(q_1^2+q_2^2)^2$ and anti Fourier transforming the result.
This gives us
$$ <T^{00}> = {{-2A_1x_1}\over {(x_1^2+x_2^2)^3}} \eqno {21} $$
for $x_1$ very much smaller than unity, $A_1$ a finite constant.

There are no infinities in this order, since the divergence of
the integral
$$ \int d^4 p {{(p_1^2+p_2^2)}\over { p^2(p-q)^2}} $$ 
has been cancelled exactly by the zero of the anti Fourier 
transforming the result.
Here we have taken
$$ \int d^2 p e^{ip_1x_1+ip_2x_2} (p_1^2+p_2^2) = 
C {{\epsilon} \over {(x_1^2+x_2^2)^2}} \eqno {22} $$
$$\int d^2 p e^ {ip_1x_1+ip_2x_2} \log (p_1^2+p_2^2) = 
C' (x_1^2+x_2^2) \log (x_1^2+x_2^2) \eqno {23} $$
                        as given in $^{/ 5  }$, and used dimensional 
                        regularization.  Here $\epsilon$ is the
                        parameter that goes to zero 
                        in the dimensional regularization and $C,C'$ are two 
   finite                     constants.

                        This example corresponds to a cosmic 
                        string at the$$x_1=-{1\over {2\alpha}} 
                        -{1\over {2 \alpha}} \sqrt{1+4 \alpha ^2},x_2=0, $$  
                        another  cosmic
                        string  with  negative of  the same  mass, 
                        located at the origin, and a
                        third cosmic string with positive mass, same as 
                        the first one located at
                       $$x_1=-{1\over {2 \alpha}} + {1\over {2 \alpha}} 
                       \sqrt {1+ 4 \alpha^2}, x_2=0. $$
                        We have in total one cosmic string and a 
                        dipole made out of cosmic strings.

                        We can calculate the vacuum fluctuation of 
                        a spinor in the presence of the same configuration.
We find that the energy of the spinor particle 
is of the same form as the scalar case.

The same formalism can be applied to other configurations, all with locally
flat Ricci scalar and Einstein tensor.  A simple extension will be 
$$
g_2=-\beta \log \left( 1+\alpha (z^2-\frac{1}{ z^2}) \right) 
$$
which    corresponds to four cosmic strings, two equally spaced 
on the $x_1$ axis 
on two sides of the
origin,  two more equally spaced on two sides of the origin on the $x_2$ axis,  
and two cosmic strings with negative mass, 
located on the origin of the $x_1-x_2$
plane.
By playing with the arbitrary functions one can get more interesting cases,
however, it is not so easy to obtain other 
than Dirac delta function type singularities
for the Ricci scalar.

   \bigskip
   \noindent
  {\bf{ Ackowledgement}}
  We thank Prof.Dr. A.N. Aliev for giving his results 
  and calculations available to us 
  prior to publication and for extensive discussions.  
  Discussions with Prof.s
  A. B\" uy\" ukaksoy, A.H. Bilge, \" O. F. Day\i 
  , Y. Nutku are also gratefully 
  acknowledged.  This work is partially supported by the the Scientific and
  Technical Research Council of Turkey and M.H.'s work 
  is also supported by the
  Academy of Sciences of Turkey.
  \vfill\eject

  \noindent
  {{\bf {REFERENCES}}
  \begin {description}

  \item {1.} T.W.B. Kibble, Phys. Reports {\bf{67}} (1980) 183, 
  M.B. Hindmarsh and
  T.W.B. Kibble, Reports  Progress Phys. {\bf{58}} (1995) 477, 
  A. Vilenkin and E.P.S. Shellard,
  {\it{ Cosmic Strings and Other Topological Defects}}, 
  Cambridge Univ. Press,
  Cambridge,  1994;

  \item {2.}  T.M. Helliwell and D.A. Konkowski, Phys. Rev. 
  {\bf{D34}} (1986) 1908;
  B. Linet, Phys. Rev {\bf{D33}} (1986) 1833, 
  Phys. Rev {\bf{D35}} (1987) 536,
  A.C.Smith in {\it{ The Formation and Evolution of 
  Cosmic Strings}}, Ed. by G.W. Gibbons,
  S.W.Hawking and T.Vachaspati, Cambridge University Press, 
  Cambridge (1990), p.263;

  \item {3.}  A.N.Aliev, " Casimir Effect in the Space-time 
  of Multiple Cosmic Strings",
  MRC preprint (1996);

  \item {4.}  P.S. Letelier, Classical and Quantum Gravity 
  {\bf{4}} (1987) L75;
   single cosmic string solutions can be found at  
   J.R. Gott III, Astrophys. J {\bf{288}}
  (1985) 442;

  \item {5.} I.M.Gelfand and G.E. Shilov, 
  {\it{ Generalized Functions}} Vol.1 p.363-364,
  Translated by E. Saletan, Academic Press, New York and London, 1964.
\end{description}
   \end {document}